\documentclass[
reprint,
amsmath,amssymb,
aps, longbibliography
]{revtex4-2}

\usepackage{amsmath, amssymb, graphicx, comment}
\usepackage{dcolumn}
\usepackage{bm, empheq}

\usepackage[dvipsnames]{xcolor}
\definecolor{links}{RGB}{0 4 245}
\usepackage[hidelinks,colorlinks=true,linkcolor=links,urlcolor=links,citecolor=links]{hyperref}

\begin{document}

\preprint{APS/123-QED}

\title{Dynamics of a nanoscale ferromagnetic vortex}

\author{Jun Seok Seo}\thanks{sjs4626@gmail.com}
\author{Se Kwon Kim}\thanks{sekwonkim@kaist.ac.kr}

\affiliation{Department of Physics, Korea Advanced Institute of Science and Technology, Daejeon 34141, Republic of Korea}

\date{\today}

\begin{abstract}
	We propose a ferromagnetic vortex stabilized by the interfacial Dzyaloshinskii--Moriya interaction (iDMI) and investigate its properties through theoretical analysis and micromagnetic simulations. Our results demonstrate that this vortex can remain stable even in nanoscale ferromagnetic disks with radii below $5\,\text{nm}$---far smaller than those of conventional nanodot vortices having about $1\,\mu\text{m}$ radius. We analytically solve the nonlinear equation of motion describing the anharmonic vortex oscillation, and identify the critical frequency that determines the stability of the driven oscillation of the vortex. This nanoscale vortex exhibits conventional properties of microscale vortices, including gyrotropic oscillation and resonance frequency shift under an out-of-plane magnetic field. It also exhibits unconventional behaviors, such as a strongly anharmonic potential, nonlinear oscillations, and a Duffing-oscillator-like response under the external AC bias.
\end{abstract}

\maketitle
\section{\label{sec:level1}INTRODUCTION}
A magnetic vortex, sometimes referred to as a meron, is a topological soliton in which the local magnetization winds around a localized vortex core~\cite{KOSEVICH1990117}. The vortex is characterized by the vortex number~\cite{Rybakov_2025},
\begin{align}
	N = \frac{1}{2\pi}\oint d\bm{\ell} \cdot \left(\frac{m_x \nabla m_y - m_y \nabla m_x}{m_x^2 + m_y^2}\right) \in \mathbb{Z}\,,
\end{align}
which is the winding number, where $\mathbf{m} = m_x \hat{\mathbf{x}} + m_y \hat{\mathbf{y}} + m_z \hat{\mathbf{z}}$ is the unit vector field representing the direction of the local magnetization. A ferromagnetic vortex can also be characterized by the Skyrmion charge~\cite{Tretiakov_2007}
\begin{align}
	\frac{pN}{2} = \frac{1}{4\pi} \int d^2 x~ \mathbf{m} \cdot \left(\partial_x \mathbf{m} \times \partial_y \mathbf{m}\right) \in \frac{\mathbb{Z}}{2}\,,
\end{align}
where $p$ indicates the two possible core polarizations: up ($p = 1$) or down ($p = -1$). Topological numbers $N$ and $p$ are integers that cannot be easily changed under continuous deformation, making them stable against slight perturbations.

Magnetic vortices have potential spintronics applications, such as in memory devices~\cite{Bohlens_2008, Pigeau_2010, Yu_2011, Hrkac_2015, 9963571}, computational devices~\cite{Torrejon_2017, Yun_2022, Shreya:2023aa, Khosla_2025}, or even biomedical technologies~\cite{Kim_2009, Rozhkova_2009}. Conventionally, a  magnetic vortex is stabilized by the demagnetization field in a ferromagnetic disk of micrometer-scale radius~\cite{doi:10.1126/science.289.5481.930}. However, this conventional stabilization mechanism relying on the magnetostatic interaction has limitations in reducing the disk size, as smaller disk may fail to support a stable vortex, which imposes a fundamental limitation on the miniaturization and integration of vortex-based applications.

\begin{figure}[t!]
	\includegraphics[width=0.85\linewidth]{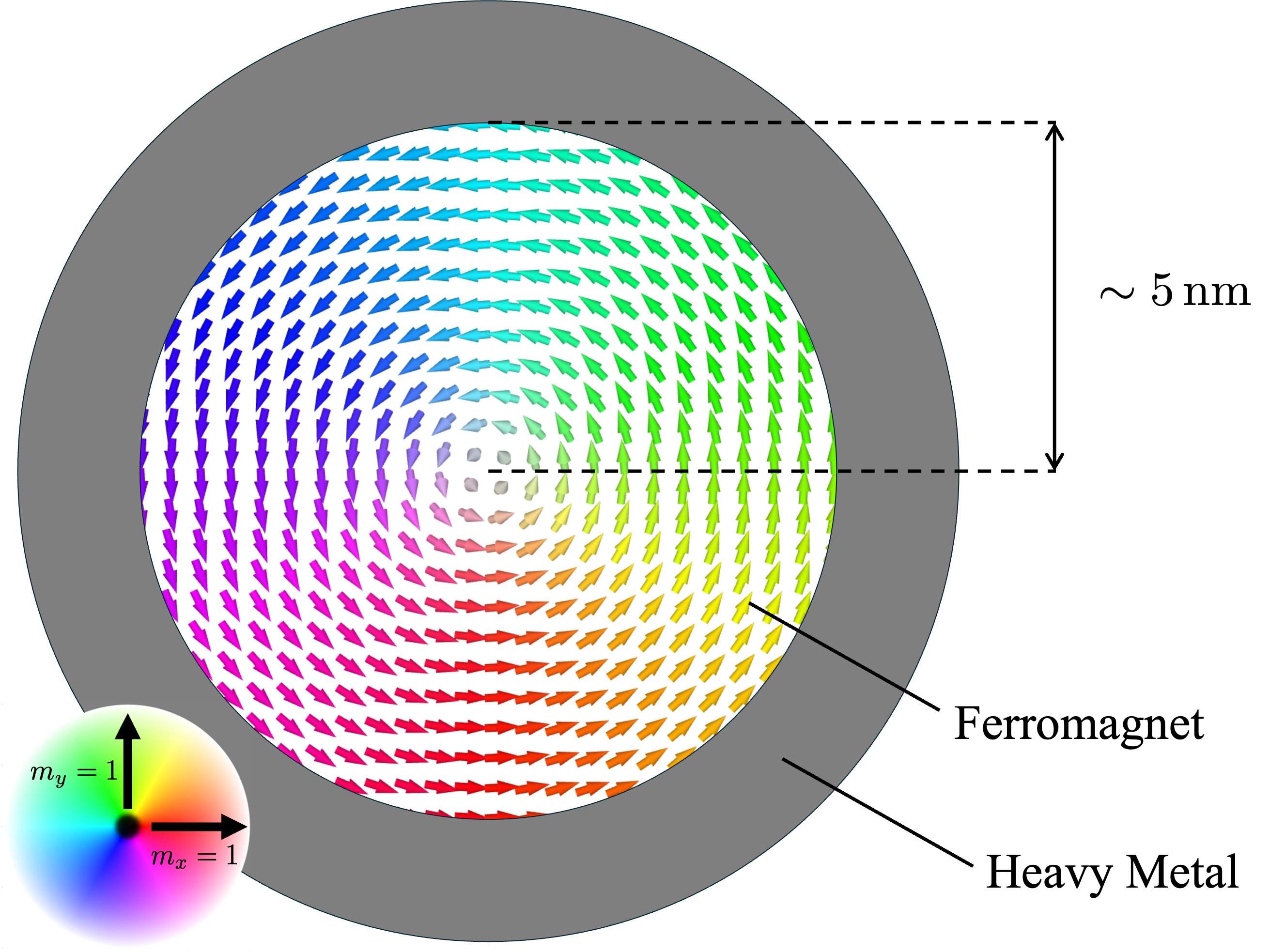}\label{fig1}
	\caption{Schematic diagram of the stabilized ferromagnetic vortex of $N = 1$, $p = 1$, in a ferromagnetic disk surrounded by heavy metal. The brightness of the spin indicates the size of the $m_z$ component: white corresponds to $m_z = +1$, and black corresponds to $m_z = -1$.}
\end{figure}

Here, we introduce a mechanism to enable a vortex soliton in a nanoscale ferromagnetic disk, utilizing the interfacial Dzyaloshinskii--Moriya interaction (iDMI)~\cite{Volkov_2018, Kuepferling_2023}. We explore the static and dynamic properties of this vortex in detail, through theoretical analysis and micromagnetic simulations using MuMax3~\cite{Vansteenkiste2014}, focusing particularly on the vortex of $N = 1$ state (see Fig.~\hyperref[fig1]{1} for the illustration). In Sec.~\hyperref[sec2]{II}, we describe the magnetic system and analyze its static properties. In Sec.~\hyperref[sec3]{III}, we investigate the vortex dynamics, emphasizing its nonlinear nature. In Sec.~\hyperref[sec4]{IV}, we examine the resonance behavior of the vortex, driven by an oscillating external bias, revealing a Duffing-oscillator like response. In Sec.~\hyperref[sec5]{V}, we provide a summary and discuss the potential applications of our nanoscale vortex. In Appendix~\hyperref[ApA]{A}, we derive the vortex potential within the anharmonic expansion. Appendix~\hyperref[ApB]{B} presents the derivation of the exact solution of the nonlinear equation of motion describing the anharmonic vortex oscillation. Appendix~\hyperref[ApC]{C} provides a detailed formulation of the driven vortex dynamics and the corresponding stability analysis.

\section{SYSTEM MODEL}\label{sec2}

A heavy metal, such as Pt, surrounding the ferromagnetic disk, induces an iDMI with an out-of-plane directional DM vector at the disk boundary, penetrating into the disk with exponentially decreasing strength~\cite{Volkov_2018, Kuepferling_2023}. We consider a ferromagnetic disk of radius $R$ and thickness $t_z$, surrounded by heavy metal. We assume that $\mathbf{m}$ is uniform along the $z$-direction, i.e., $\mathbf{m}(x, y, z) = \mathbf{m}(x, y)$. Let $C$ denote the domain of the disk. The energy functional of the continuum model describing the system is given by
\begin{equation}\label{energy}
	\begin{gathered}
		U[\mathbf{m}] = \frac{J}{2} \int_C d^2 x ~ (\nabla \mathbf{m})^2 + \frac{K}{2} \int_C d^2 x~ (\hat{\mathbf{z}} \cdot \mathbf{m})^2\\
		+ U_{\text{iDMI}}[\mathbf{m}]  + U_\text{demag}[\mathbf{m}]\,.
	\end{gathered}
\end{equation}
Here, $J > 0$ is the exchange stiffness and $K > 0$ is the anisotropy constant. Interaction between the ferromagnet and the surrounding heavy metal induces $U_\text{iDMI}$, which can be expressed as
\begin{align}\label{iDMI}
	- \frac{D}{\delta_0 (1 - e^{-R/\delta_0})} \int_{C} d^2 x~ e^{-\frac{R - r}{\delta_0}}\hat{\mathbf{z}} \cdot \left(\mathbf{m} \times \frac{1}{r}\partial_\varphi \mathbf{m}\right),
\end{align}
where $D$ is the iDMI strength and $\delta_0$ is the penetration depth of iDMI into the boundary. Under the small penetration depth limit $\delta_0 \to 0$, Eq.~\eqref{iDMI} becomes
\begin{gather}
	-D \oint_{\partial C} d\ell\ \hat{\mathbf{z}} \cdot \left(\mathbf{m} \times \frac{1}{r} \partial_\varphi \mathbf{m}\right),
\end{gather}
which is the conventional iDMI energy functional~\cite{Volkov_2018}.

\begin{figure}[t!]
	\includegraphics[width=\linewidth]{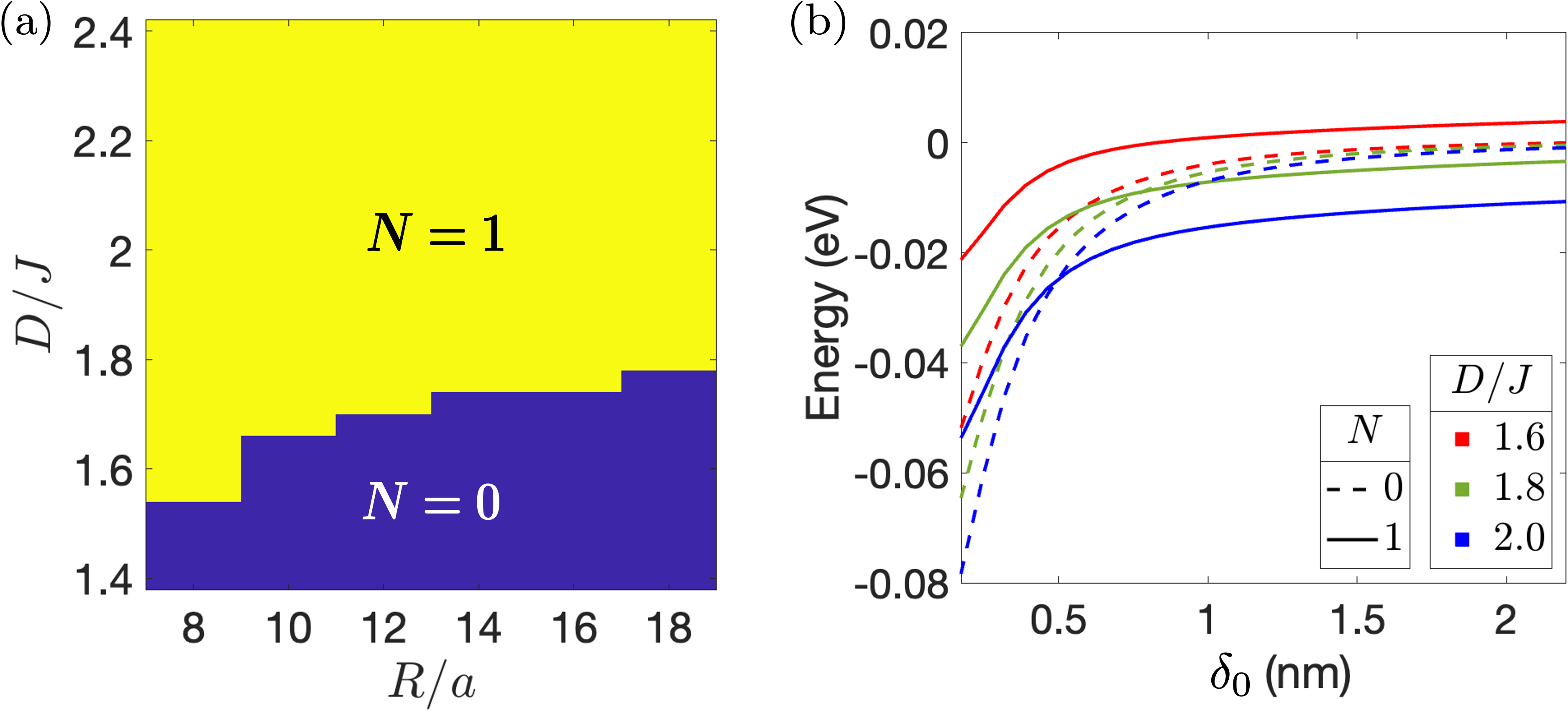}\label{fig2}
	\caption{(a) $R/a$ versus $D/J$ phase diagram of the vortex number $N$, obtained by comparing the energy Eq.~\eqref{energy} of the relaxed state from $N = 0 \sim 3$ with fixed $J = 1\times 10^{-21}\,\text{J}$. (b) Comparison between the energy Eq.~\eqref{energy} of the relaxed states with $N = 0$ and $N = 1$ for various iDMI strengths. Dotted lines and solid lines correspond to $N = 0$ and $N = 1$, respectively, while different colors denote different $D / J$ ratios at a fixed $J = 1\times10^{-21}\,\text{J}$.}
\end{figure}

Our system is a nanoscale disk (with both radii and thickness of a few nanometers) and therefore its static and dynamic properties of spin textures can be well described by considering the short-range terms---the exchange interacation, the magnetic anisotropy, and the iDMI---with little influence from the the long-range interaction $U_\text{demag}$ term.~\cite{Guslienko_2001}. Conventional ferromagnetic nanodot vortex necessarily requires a sufficiently large demagnetization field, competing with the exchange interaction, to stabilize the vortex. Instead, the relatively simple local iDMI term engenders the vortex structure, replacing the role of the complex nonlocal $U_\text{demag}$ term of the conventional nanodot vortices. Our analytical theory in this work is based on the short-range terms, but the micromagnetic simulations are performed with the demagnetizing field included. The agreement between the two that shall be discussed below shows that the effect of the demagnetization field is indeed insignificant for nanoscale vortices.

Throughout this paper, Eq.~\eqref{iDMI} is implemented in the simulations, using the custom field feature of MuMax3~\cite{Zou_2025}. Unless otherwise stated, we fix $N = 1$, $p = 1$, $h = \pi / 2$, $a = t_z = 0.4\,\text{nm}$ where $a$ is a lattice constant of the simple cubic lattice, $R = 4.8\,\text{nm}$, $\delta_0 = 2.0\,\text{nm}$, $D = 2 \times 10^{-21}\,\text{J}$, $A_\text{ex} = J / (2a)=1.25\times10^{-12}\,\text{J/m}$, $K_{u1} = -2aK = -8\times10^{-4}\,\text{J/m}^3$, and $M_\text{sat} = 500 \times 10^3\,\text{A/m}$ for a simulation~\cite{Coey_2010, Kowacz_2022}.

Since the exchange interaction favors a uniform spin configuration, the vortex tends to be expelled from a finite-sized disk, to evolve into a uniform spin state. On the other hand, the iDMI with the out-of-plane DM vector favors chiral spin rotation along the disk boundary, whereby promoting the vortex spin state. The competition between the exchange interaction and the iDMI determines whether the vortex can be stabilized~\cite{Zou_2025}. Figure \hyperref[fig1]{1} shows a schematic diagram of the stabilized vortex under iDMI, in a ferromagnetic disk surrounded by heavy metal. The phase diagram of $N$, showing $R/a$ versus $D / J$, is presented in Fig.~\hyperref[fig2]{2(a)}, obtained from simulations. The vortex tends to stabilize more easily with smaller $R$ and larger $D$. In addition, the energy comparison by simulation, between the $N = 0$ and $N = 1$ states, is shown in Fig.~\hyperref[fig2]{2(b)}. A sufficiently large $\delta_0$ can stabilize the vortex state ($N = 1$); however, if $\delta_0$ is not large enough, the system more favors the parallel spin configuration ($N = 0$). The relaxation process of the vortex of $N = 1$ is shown in Supplemental Video 1~\cite{suppl}.

Denoting $\mathbf{m}(\mathbf{r}) = (\cos\phi \sin\theta, \sin\phi \sin\theta, \cos\theta)$ and $\mathbf{r} = (r \cos\varphi, r\sin\varphi)$, the ansatz for the vortex configuration we adopt is
\begin{gather}
	\theta(r, \varphi) = -\frac{\pi}{2}p + 2\arctan\left[\exp\left(-\beta\frac{r}{\lambda_0}\right)\right],\label{ansatz_theta}\\
	\phi(r, \varphi) = \varphi + h~,\label{ansatz_phi}
\end{gather}
where $\lambda_0 \equiv \sqrt{J/K}$ is the characteristic core size of the vortex. For convenience, we define the dimensionless parameters $\epsilon \equiv R / \lambda_0$ and $\mu \equiv R / \delta_0$. The variational parameter $\beta$ is determined by minimizing the energy functional $E(\beta) \equiv U[\mathbf{m}]$~\cite{PhysRevResearch.1.033109}, i.e., a $\beta$ which satisfies $dE/d\beta = 0$ \footnote{Explicitly,
\begin{gather*}
	\frac{1}{2\pi}\frac{dE}{d\beta}	=\frac{J}{2}\Biggr[\left(\beta \epsilon^2 + \frac{\epsilon^2 - 1}{\beta}\right)\text{sech}^2(\beta\epsilon)\\
	+ \frac{2}{\beta^3}\ln\left[\cosh(\beta\epsilon)\right] - \frac{2\epsilon}{\beta^2}\tanh(\beta\epsilon) + \frac{1}{\beta}\Biggr] \\
		- \frac{2}{\beta^2}\frac{D \mu}{\epsilon \left(e^\mu - 1\right)} \int_{0}^{\beta \epsilon} d\tau~\tau \exp\left(\frac{\mu}{\beta \epsilon}\tau\right) \tanh \tau\, \text{sech}^2 \tau\,.
\end{gather*}
}
except the trivial solution $\beta = 0$; the equation can be solved numerically. For example, under our simulation conditions, $\beta \approx 1.40$. The helicity $h$ of the vortex is chosen as $h = \pm \pi / 2$, due to the $U(1)$ symmetry breaking induced by the magnetostatic interaction~\cite{PYLYPOVSKYI2014201}.

When the vortex core is slightly displaced from the disk center by $\bm{\mathcal{R}} = \mathcal{X}\hat{\mathbf{x}} +  \mathcal{Y}\hat{\mathbf{y}}$, the corresponding potential for a vortex state with $N = 1$ is given by $V(\mathcal{X}, \mathcal{Y}) \equiv U[\mathbf{m}(\mathbf{r} - \bm{\mathcal{R}})]$. Due to the system's rotational symmetry, the potential is isotropic, hence it can be expanded as:
\begin{align}\label{potential}
	V(\mathcal{R}) = V_0 + \frac{1}{2} c_2 \mathcal{R}^2 + \frac{1}{4!} c_4 \mathcal{R}^4 + \cdots.
\end{align}
The coefficients $c_{2k}$ for $k = 1, 2, 3,\cdots$ can be calculated exactly and are given by
\begin{align}\label{cn}
	c_{2k} &\,=\frac{2\pi}{\lambda_0^{2k}}\left[D f^{(2k)}(\beta, \epsilon, \mu) - J g^{(2k)}(\beta, \epsilon)\right],
\end{align}
where $f^{(2k)}$ and $g^{(2k)}$ are dimensionless functions, whose explicit expressions and derivations can be found in Appendix~\hyperref[ApA]{A}.

To stabilize the vortex at the center of the disk, the condition $c_2 > 0$ must be satisfied; otherwise, the vortex will be expelled from the center. However, the exact condition for the stability of the vortex depends on more than just the sign of $c_2$; a comparison between the total energy between states with different $N$ should be considered.

\begin{figure}[t!]
	\includegraphics[width=\linewidth]{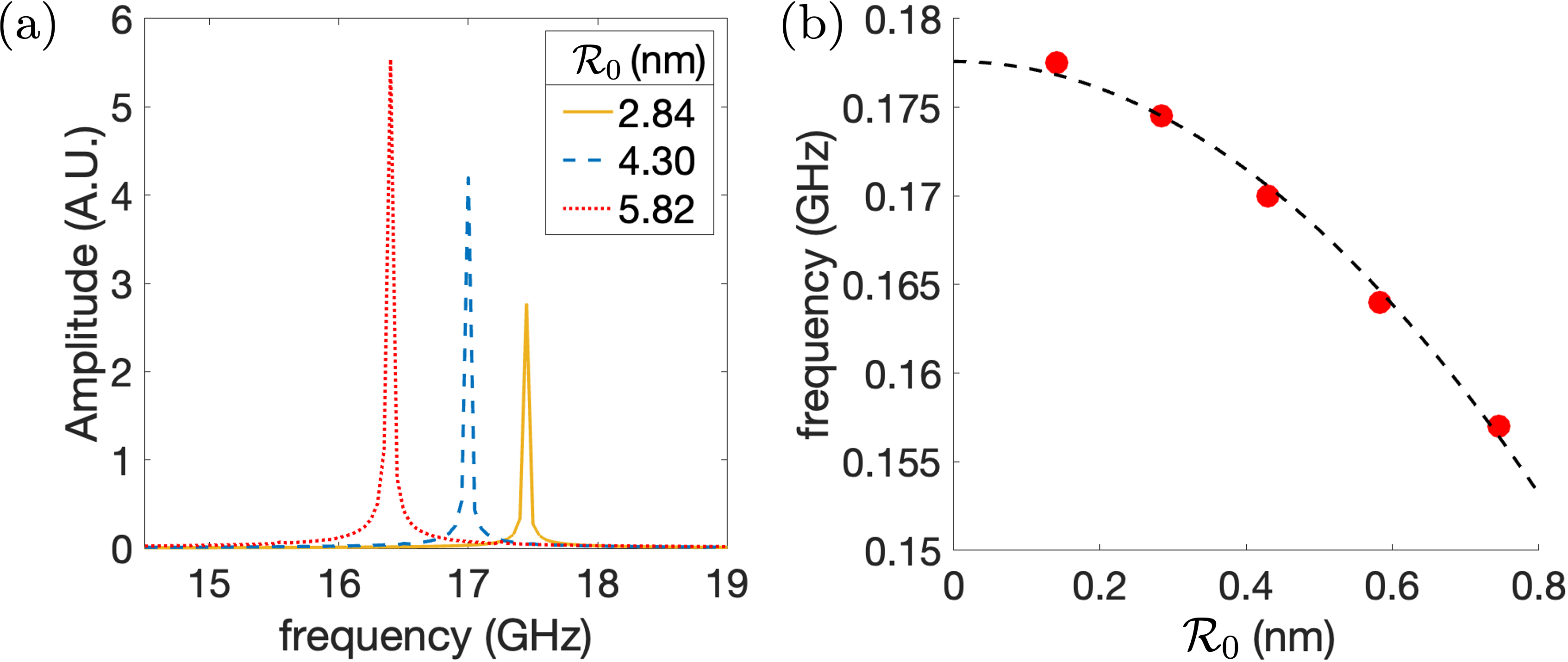}\label{fig3}
	\caption{(a) Simulation results of the amplitude-frequency response for various $\mathcal{R}_0$ at $\alpha = 0$. (b) Relation between $\mathcal{R}_0$ and the oscillation frequency at $\alpha = 0$, obtained from simulations (red dots) and fitted to Eq.~\eqref{omega_0} divided by $2\pi$ (dashed line).}
\end{figure}

A relevant length scale characterizing the importance of the quartic term is $\xi \equiv \sqrt{|6c_2 / c_4|}$, since
\begin{equation}\label{potential2}
\begin{aligned}
	V(\mathcal{R}) &\,\approx V_0 + \frac{1}{2} c_2 \mathcal{R}^2 + \frac{1}{4!} c_4 \mathcal{R}^4\\
	&\,= V_0 + \frac{1}{2} c_2 \mathcal{R}^2 \left(1 \pm \frac{\mathcal{R}^2}{\xi^2}\right).
\end{aligned}
\end{equation}
If $\xi$ is sufficiently larger than the small displacement $\mathcal{R}$, the quartic potential can be neglected, and then the oscillation can be regarded as harmonic. However, using Eq.~\eqref{cn}, our simulation condition yields $c_2 = 9.68 \times 10^{-4}\,\text{J/m}^2$ and $c_4 = -1.05\times 10^{15}\,\text{J/m}^4$, so that $\xi \sim 6 a$, the quartic term is difficult to be neglected even for the small displacements about $a$ of the vortex core, and significantly affects the vortex dynamics~\cite{Guslienko_2010, Guslienko:2014aa, Drews_2012, Cheng_2009}. Also, using Eq.~\eqref{cn}, we numerically confirmed that for the parameters satisfying the necessary condition for vortex stability (i.e., $c_2 > 0$), $\xi$ is small and only weakly dependent on $D$, $R$, and $\delta_0$. This indicates that the anharmonicity of our nanoscale vortex cannot be easily neglected.

\begin{figure}[t!]
	\includegraphics[width=\linewidth]{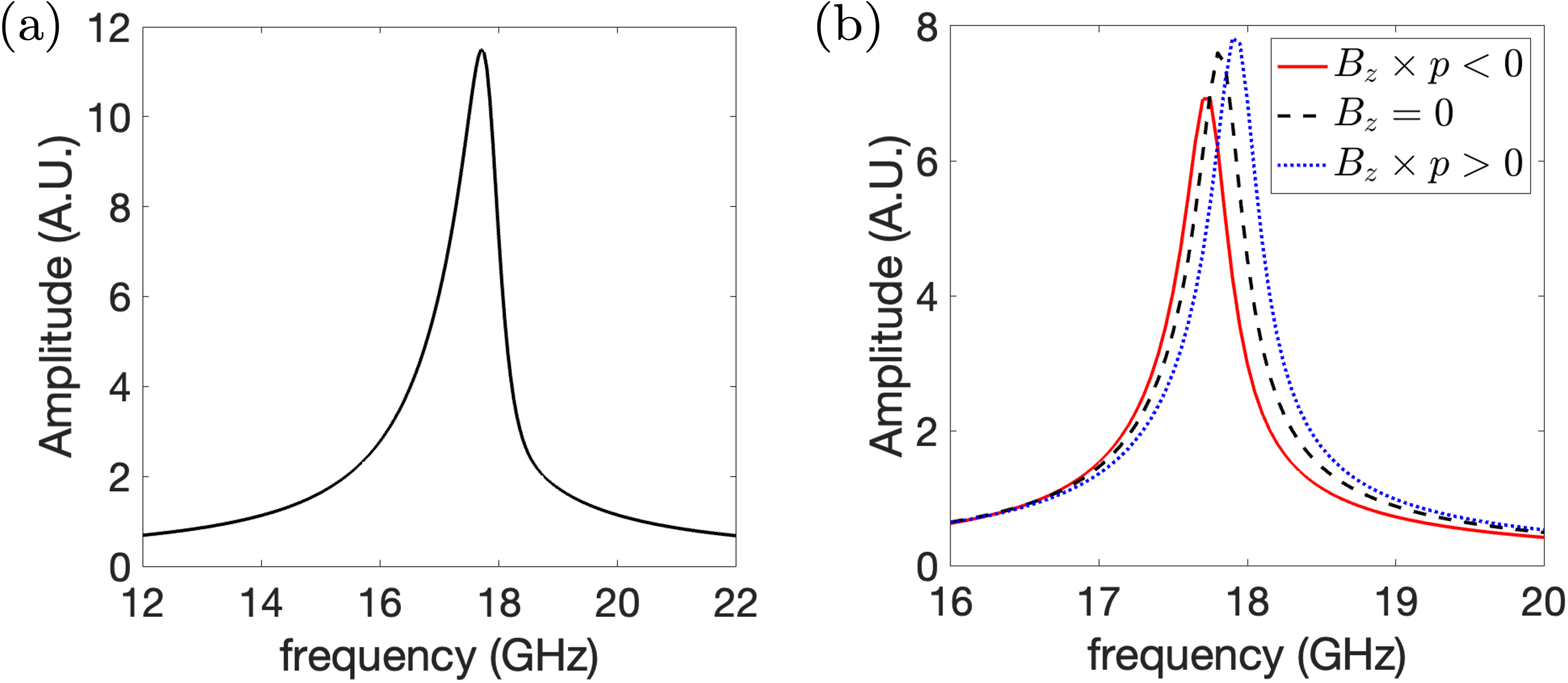}\label{fig4}
	\caption{(a) Simulation result of the amplitude-frequency response of the vortex oscillation, for $\mathcal{R}_0 = 0.37\,\text{nm}$ and $\alpha = 5\times10^{-3}$. (b) Shift of the oscillation frequency under $\mathbf{B} = B_z \hat{\mathbf{z}}$, simulation at $\mathcal{R}_0 = 0.12\,\text{nm}$ and $\alpha = 5\times10^{-3}$. Blue (red) curve shows that oscillation frequency at $B_z = 120\,\text{mT}$ ($-120\,\text{mT}$), which increases (decreases) as $B_z$ is in same (different) direction with polarity of the vortex, from the oscillation frequency with no external field (dotted line).}
\end{figure}

\section{DYNAMICS OF THE VORTEX}\label{sec3}
A slight perturbation of the vortex core leads to gyrotropic oscillation of the vortex~\cite{Guslienko_2002, doi:10.1126/science.1095068}. Thiele's equation of motion~\cite{PhysRevLett.30.230}---which is based on the Landau-Lifshitz-Gilbert equation---for a vortex of $N = 1$~\cite{Guslienko_2002, Choi_2009, Hrkac_2015}, under the potential in Eq.~\eqref{potential} expanded up to quartic order in $\mathcal{R}$, is
\begin{equation}\label{thiele}
\begin{aligned}
	&\,\begin{pmatrix}
		\alpha I&-p\\
		p&\alpha I
	\end{pmatrix}
	\begin{pmatrix}
		\dot{\mathcal{X}}\\
		\dot{\mathcal{Y}}
	\end{pmatrix}
	=
	-\frac{\gamma \left(c_2 + \frac{c_4}{6}\mathcal{R}^2\right)}{2\pi  M_\text{sat} t_z}
	\begin{pmatrix}
		\mathcal{X}\\
		\mathcal{Y}
	\end{pmatrix},
\end{aligned}
\end{equation}
where $\gamma$ is the gyromagnetic ratio of the spin, and $I = I(\beta, \epsilon) \equiv \int_C d^2x~ (\partial_x \mathbf{m})^2 = \int_C d^2x~ (\partial_y \mathbf{m})^2$, which is given by
\begin{equation}
	\begin{aligned}
		I\!=\!\frac{1}{2}\!\left[\beta \epsilon \tanh(\beta  \epsilon) \!-\! \ln\left[\cosh(\beta\epsilon)\right]\!+ \!\!\int_{0}^{\beta \epsilon}\!\!\! d\tau\, \frac{\tanh^2 \tau}{\tau}\!\right]
	\end{aligned}.
\end{equation}
Defining $\psi \equiv \mathcal{X} + i\mathcal{Y}$, Eq.~\eqref{thiele} can be rewritten as
\begin{equation}\label{ode}
	\begin{aligned}
		\dot{\psi} = -\frac{\gamma \left(c_2 + \frac{c_4}{6}|\psi|^2\right)\psi}{2\pi M_\text{sat} t_z \left[\alpha I(\beta, \epsilon) + i p\right]} \equiv i\omega \psi + i \lambda |\psi|^2 \psi\,.
	\end{aligned}
\end{equation}

One simple case arises when $\alpha = 0$, in which case the radius of gyration $\mathcal{R} = |\psi|$ remains constant as $\mathcal{R}_0 \equiv |\psi_0|$, where $\psi_0 \equiv \psi(t = 0)$~\cite{Guslienko:2014aa, Guslienko_2010, Cheng_2009}. In this case, Eq.~\eqref{ode} reduces to the equation of a harmonic oscillator $\dot{\psi} = i \left(\omega + \lambda \mathcal{R}_0^2\right)\psi$, whose frequency is
\begin{gather}\label{omega_0}
	\omega_{\alpha = 0}(\mathcal{R}_0) \equiv \omega_r + \lambda_r \mathcal{R}_0^2\,,
\end{gather}
where $\omega_r$ and $\lambda_r$ denote the real parts of $\omega$ and $\lambda$, respectively. Figure \hyperref[fig2]{3(a)} shows the amplitude-frequency response for various $\mathcal{R}_0$ at $\alpha = 0$, obtained from simulations. The response peaks are symmetric and the oscillation frequency decreases as $\mathcal{R}_0$ increases. Fig.~\hyperref[fig2]{3(b)} presents the relation between $\mathcal{R}_0$ and oscillation frequency at $\alpha = 0$, obtained from simulation and fitted to Eq.~\eqref{omega_0} divided by $2\pi$, to extract $\omega_r$ and $\lambda_r$. From fitting the simulation results, we obtain $\omega_r \approx 1.12\times10^{11}\,\text{s}^{-1}$ and $\lambda_r \approx -2.39\times 10^{28}\,\text{m}^{-2}\cdot\text{s}^{-1}$, which are in good agreement with the theoretical values $\omega_r \approx 1.36\times10^{11}\,\text{s}^{-1}$ and $\lambda_r \approx -2.45\times 10^{28}\,\text{m}^{-2}\cdot\text{s}^{-1}$, computed by Eq.~\eqref{omega_0}.

If $\alpha > 0$, the radius of gyration gradually decreases over time, and the anharmonicity emerges. Supplemental Video 2 shows the vortex oscillation, comparing the cases with $\alpha = 10^{-2}$ and $\alpha = 0$~\cite{suppl}. Equation \eqref{ode} can also be solved exactly in this general case:
\begin{align}\label{sol}
	\psi(t) = \psi_0 \exp\left[i \omega t + i \lambda \sigma(t)\right]\,,
\end{align}
where
\begin{align}
	\sigma(t) \equiv \frac{1}{\lambda_i} \ln \sqrt{1 + \frac{\lambda_i \mathcal{R}_0^2}{\omega_i}\left(1 - e^{-2\omega_i t}\right)}\,,
\end{align}
with $\omega_i$ and $\lambda_i$ denoting the imaginary parts of $\omega$ and $\lambda$, respectively. See Appendix~\hyperref[ApB]{B} for a detailed derivation.

In contrast to the harmonic oscillation, this anharmonic oscillation shows an asymmetric amplitude-frequency response as shown in Fig.~\hyperref[fig4]{4(a)}. Amplitude-frequency responses are obtained from simulation, which is a modulus of the finite fourier transform of $\psi(t)$. 

Under the out-of-plane external magnetic field $\mathbf{B} = B_z \hat{\mathbf{z}}$, oscillation frequency of the vortex varies as the gyrotropic tensor component of the Thiele's equation changes~\cite{de_Loubens_2009}. Fig.~\hyperref[fig4]{4(b)} shows the simulation result with $\alpha = 5\times 10^{-3}$ and $\mathcal{R}_0 = 0.12\,\text{nm}$, under the external magnetic field $B_z = -120\,\text{mT}$, $0\,\text{mT}$, and $+120\,\text{mT}$. The amplitude-frequency response is asymmetric, but we can observe a shift of the curve under the external magnetic field.

\begin{figure}[t!]
	\includegraphics[width=\linewidth]{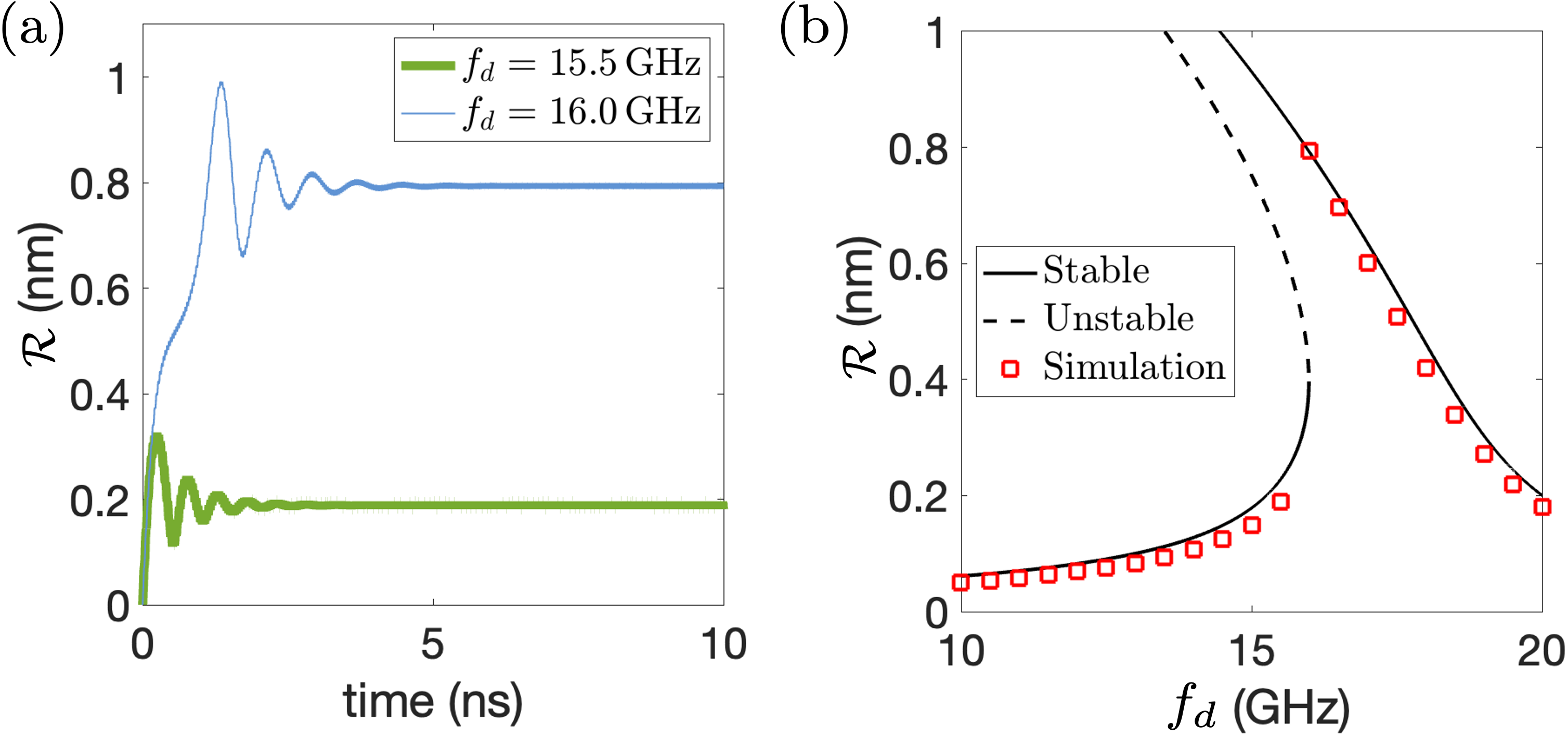}\label{fig5}
	\caption{(a) Stabilization of radius of gyration at driving frequencies $f_d = 15.5\,\text{GHz}$ (green, thick) and $f_d = 16.0\,\text{GHz}$ (blue, thin). The gyration radius stabilizes over time. (b) Comparison between the numerically obtained frequency-amplitude curve (black line) and the simulation results for the time-averaged quasi-stationary radius of gyration (red squares). Solid (dashed) lines indicate dynamically stable (unstable) solutions.}
\end{figure}

\section{RESONANCE BEHAVIOR OF THE DRIVEN OSCILLATION}\label{sec4}
When an oscillatory external torque---such as an external magnetic field $\mathbf{B}(t) = B_x \sin(\omega_d t)\hat{\mathbf{x}}$ or the adiabatic Zhang--Li STT~\cite{Li_2004} with $\mathbf{j}(t) = j_x \sin(\omega_d t) \hat{\mathbf{x}}$---is applied, a conventional vortex with linear dynamics exhibits oscillations with a certain quasi-stationary radius of gyration, reaching a maximum when the vortex oscillation frequency matches the driving frequency~\cite{Novosad_2005, Kasai_2006}. However, nonlinear vortex dynamics can exhibit a different responses under driving oscillation~\cite{Guslienko_2010, Drews_2012}.

The equation of motion under an oscillatory external torque is written as
\begin{align}\label{driven}
	\dot{\psi} = i\omega \psi + i\lambda |\psi|^2\psi + F_0 \sin(\omega_d t)\,,
\end{align}
where $F_0$ is a constant proportional to either $B_x$ or $j_x$~\cite{suppl}. The frequency-amplitude response of this driven oscillation can be obtained using the method of the harmonic balance~\cite{10.1093/oso/9780199208241.001.0001}, yielding
\begin{equation}
	\begin{aligned}\label{response}
		|\omega_d - \left(\omega + \lambda \mathcal{R}^2\right)|^2 \mathcal{R}^2 = \frac{|F_0|^2}{2},
	\end{aligned}
\end{equation}
and dynamical stability of these solutions is determined via Floquet-Lyapunov theory (see Appendix~\hyperref[ApC]{C} for details).

We performed micromagnetic simulations for $\alpha = 10^{-2}$, applying an oscillating external magnetic field with a fixed amplitude of $B_x = 10\,\text{mT}$ and various driving frequencies $f_d \equiv \omega_d / (2\pi)$. The radius of gyration relaxes to quasi-stationary value over time under the applied torque (Fig.~\hyperref[fig5]{5(a)}). The red squares in Fig.~\hyperref[fig5]{5(b)} represent the time-averaged quasi-stationary radius of gyration obtained from simulations for various driving frequencies. In Fig.~\hyperref[fig5]{5(b)} the black line plots the numerically obtained solutions of Eq.~\eqref{response}, where the theoretical values of $\omega$ and $\lambda$ were adjusted by applying a scaling factor based on the simulation result in Sec.~\hyperref[sec3]{III}. The dynamically stable (unstable) branches are indicated by solid (dashed) lines. As shown in Fig.~\hyperref[fig5]{5(b)}, the response amplitude exhibits a jump at a certain driving frequency. The simulation results show close agreement with the theoretical prediction and explains the abrupt jump in quasi-stationary amplitude occuring between $f_d = 15.5\,\text{GHz}$ and $f_d = 16.0\,\text{GHz}$. This frequency-amplitude response is analogous to that of a driven Duffing oscillator~\cite{10.1093/oso/9780199208241.001.0001}.

\section{DISCUSSION AND CONCLUSIONS}\label{sec5}
In this paper, we proposed a ferromagnetic vortex stabilized by the iDMI and demonstrated its properties through theoretical modeling and micromagnetic simulations. Unlike conventional vortices stabilized by demagnetization energy, our system enables stable vortex formation in nanoscale disks with radii on the order of nanometers, due to penetrating interfacial effects. This vortex retains conventional behaviors such as frequency shifts under out-of-plane magnetic fields, while also exhibits nontrivial features. Due to the strong anharmonicity in the potential, the vortex exhibits nonlinear dynamics, and especially the driven dynamics is similar to those of a Duffing oscillator. These behaviors are well described by analytical calculations and shows good agreement with simulation results. In particular, the good agreement between the theoretical and simulated critical behavior of the driven oscillation suggests that our findings may be useful for future experimental realizations~\cite{Novosad_2005, Kasai_2006}.

A magnetic vortex has potential in applications, such as memory devices, nano-oscillators, or neuromorphic computing. This nanoscale vortex, due to its smaller size and faster core switching dynamics, holds promise for the development of enhanced spintronic devices. Moreover, the nanoscale size of the vortex suggests the possibility of entering the quantum regime, potential in exhibiting quantum phenomena such as quantum tunneling of helicity~\cite{PhysRevB.106.104422} or quantum effects in vortex--vortex interactions~\cite{PhysRevB.110.104411}. The possibility of such quantum phenomena indicates that our nanoscale magnetic vortex may serve as a potential platform for quantum computing elements, such as qubits. Further studies are expected on vortex stabilization at room temperature, vortex-vortex interaction~\cite{PhysRevLett.106.197203}, and behavior in the quantum regime.

\begin{acknowledgments}
	J.S.S. and S.K.K. were supported by Brain Pool Plus Program through the National Research Foundation of Korea funded by the Ministry of Science and ICT (2020H1D3A2A03099291) and Basic Science Research Program through the National Research Foundation of Korea (NRF) funded by the Ministry of Education (2019R1A6A1A10073887).
\end{acknowledgments}

\onecolumngrid
\appendix
\noindent\rule{\linewidth}{0.6pt}

\section{Evaluation of the vortex potential}\label{ApA}
In this section, we evaluate the vortex potential $V(\mathcal{R})$, as defined in Eq.~\eqref{potential}. Let $u$ denote the energy density, i.e., $U[\mathbf{m}] = \int_C d^2x~u[\mathbf{m}]$. By the definition of $V$,
\begin{equation}
	\begin{aligned}\label{eq1}
		&\,V(\mathcal{X}, \mathcal{Y}) = U[\mathbf{m}(x - \mathcal{X}, y - \mathcal{Y})] = \int_C d^2x~ u[\mathbf{m}(x - \mathcal{X}, y - \mathcal{Y})]\\
		&\,=\int_C d^2x~\left(u + \left[(\partial_x u) \mathcal{X} + (\partial_y u) \mathcal{Y}\right] + \frac{1}{2}\left[(\partial_x^2 u)\mathcal{X}^2 + 2(\partial_x \partial_y u)\mathcal{X}\mathcal{Y} + (\partial_y^2 u)\mathcal{Y}^2\right] + \cdots\right).
	\end{aligned}
\end{equation}
By comparing Eq.~\eqref{eq1} with the readily-known expansion $V(\mathcal{R}) = \frac{1}{2}c_2 \mathcal{R}^2 + \frac{1}{4!}c_4 \mathcal{R}^4 + \cdots$ from the symmetry, we conclude that
\begin{gather}
	\int_C d^2x~\partial_\mathcal{X} u\big|_{\mathcal{R} = 0} = \int_C d^2x~\partial_\mathcal{Y} u\big|_{\mathcal{R} = 0} = 0\,,\\
	\int_C d^2x~\partial_\mathcal{X}^2 u\big|_{\mathcal{R} = 0} = \int_C d^2x~\partial_\mathcal{Y}^2 u\big|_{\mathcal{R} = 0} = c_2\quad\text{and}\quad\int_C d^2 x~\partial_\mathcal{X}\partial_\mathcal{Y} u\big|_{\mathcal{R} = 0} = 0\,,
\end{gather}
$\cdots$, and so on. Hence, $c_2 = \frac{1}{2}\int_C d^2x~\left(\partial_\mathcal{X}^2 + \partial_\mathcal{Y}^2\right)u = \frac{1}{2}\int_C d^2x~\nabla_{\bm{\mathcal{R}}}^2 u|_{\mathcal{R} = 0}\,$. In general, we find that
\begin{align}
	c_{2k} = \frac{1}{2^k}\int_C d^2x~\left(\nabla_{\bm{\mathcal{R}}}^2\right)^k u\big|_{\mathcal{R} = 0}\qquad(k = 1, 2, 3, \cdots)\,.
\end{align}
Note that the energy density $u$ may depend explicitly on position $\mathbf{r}$, independently of the configuration $\mathbf{m}(\mathbf{r})$. For clarity, we write $u = u[\tilde{\mathbf{r}}, \mathbf{m}(\mathbf{r})]$, where $\tilde{\mathbf{r}}$ is the intrinsic spatial dependence of $u$, so that the translation about $\bm{\mathcal{R}}$ of the spin configuration will result in $u[\tilde{\mathbf{r}}, \mathbf{m}(\mathbf{r} - \bm{\mathcal{R}})]|_{\tilde{\mathbf{r}} = \mathbf{r}}$. Hence
\begin{align}
	c_{2k} = \frac{1}{2^k}\int_C d^2x~\left(\nabla_{\bm{\mathcal{R}}}^2\right)^k u[\tilde{\mathbf{r}}, \mathbf{m}(\mathbf{r} - \bm{\mathcal{R}})]\big|_{\tilde{\mathbf{r}} = \mathbf{r}, \mathcal{R} = 0} = \frac{1}{2^k} \int_C d^2x~ \left(\nabla_\mathbf{r}^2\right)^k u[\tilde{\mathbf{r}}, \mathbf{m}(\mathbf{r})]\big|_{\tilde{\mathbf{r}} = \mathbf{r}}\,.
\end{align}
The exchange, anisotropy, and iDMI energy densities are each given by
\begin{align}
	u_{\text{ex}}[\tilde{\mathbf{r}}, \mathbf{m}(\mathbf{r})] &\,= \frac{J}{2}\left([\theta^\prime(r)]^2 + \left[\frac{\sin\theta(r)}{r}\right]^2\right),\\
	u_{\text{anis}}[\tilde{\mathbf{r}}, \mathbf{m}(\mathbf{r})] &\,= \frac{K}{2}\cos^2\theta(r)= \frac{J}{2}\frac{1}{\lambda_0^2}\cos^2\theta(r)\,,\\
	u_\text{iDMI}[\tilde{\mathbf{r}}, \mathbf{m}(\mathbf{r})] &\,= -\frac{D}{\delta_0[1 - \exp(-R / \delta_0)]} e^{-\frac{R - \tilde{r}}{\delta_0}}\hat{\mathbf{z}} \cdot \left(\mathbf{m} \cdot (\hat{\tilde{\bm{\varphi}}} \cdot \nabla_\mathbf{r} \mathbf{m})\right)= -\frac{D}{\delta_0\left(e^\mu - 1\right)} \exp(\tilde{r}/\delta_0)\frac{\sin^2\theta(r)}{r} \cos(\varphi - \tilde{\varphi})\,.
\end{align}
Using the ansatz Eq.~\eqref{ansatz_theta} and \eqref{ansatz_phi},
\begin{align}
	u_\text{ex}(\tilde{\mathbf{r}}, \mathbf{r}) &\,= \frac{J}{2}\left[\frac{\beta^2}{\lambda_0^2}\,\text{sech}^2\left(\frac{\beta}{\lambda_0}r\right) + \frac{1}{r^2}\tanh^2\left(\frac{\beta}{\lambda_0}r\right)\right],\\
	u_\text{anis}(\tilde{\mathbf{r}}, \mathbf{r}) &\,= \frac{J}{2}\,\frac{1}{\lambda_0^2}\text{sech}^2\left(\frac{\beta}{\lambda_0}r\right)\,,\\
	u_\text{iDMI}(\tilde{\mathbf{r}}, \mathbf{r}) &\,= -\frac{D}{\delta_0\left(e^{\mu} - 1\right)} \exp(\tilde{r}/\delta_0) \frac{1}{r}\tanh^2\left(\frac{\beta}{\lambda_0}r\right) \cos(\varphi - \tilde{\varphi})\,.
\end{align}
Since $u_\text{ex} (\tilde{\mathbf{r}}, \mathbf{r}) = u_\text{ex}(r)$ and $u_\text{anis} (\tilde{\mathbf{r}}, \mathbf{r}) = u_\text{anis}(r)$,
\begin{equation}
	\begin{aligned}
		&\,\int_C d^2x~ \left(\nabla_\mathbf{r}^2\right)^k \left(u_\text{ex} + u_\text{anis}\right)\big|_{\tilde{\mathbf{r}} = \mathbf{r}} = 2\pi \int_{0}^{R} dr~r\left[\frac{1}{r}\partial_r \left(r \partial_r\right)\right]^k \left(u_\text{ex} + u_\text{anis}\right)\\
		&\,= J \frac{2\pi}{\lambda_0^{2k}}\frac{\beta^{2k - 2}}{2} \left(\tau \partial_\tau \left[\frac{1}{\tau}\partial_\tau(\tau\partial_\tau)\right]^{k - 1}\left((\beta^2 + 1)\,\text{sech}^2\tau + \frac{\beta^2}{\tau^2} \tanh^2\tau\right)\right)\biggr|_{\tau = 0}^{\beta \epsilon}.
	\end{aligned}
\end{equation}
$u_\text{iDMI}$ cannot be evaluated in the same way, as it depends on both $\tilde{\mathbf{r}}$ and $\mathbf{r}$. However, since $(\nabla_\mathbf{r}^2)^k u_\text{iDMI}|_{\tilde{\mathbf{r}} = \mathbf{r}}$ is independent of $\varphi$,
\begin{equation}
	\begin{aligned}
		&\,\int_C d^2x~ \left(\nabla_\mathbf{r}^2\right)^k u_\text{iDMI} \big|_{\tilde{\mathbf{r}} = \mathbf{r}} = 2\pi \int_{0}^{R} dr~ r\left(\frac{1}{r}\partial_r(r\partial_r) + \frac{1}{r^2}\partial_\varphi^2\right)^k u_\text{iDMI}\big|_{\tilde{\mathbf{r}} = \mathbf{r}}\\
		&\,= - D \frac{2\pi}{\lambda_0^{2k}} \frac{\beta^{2k - 1} \mu}{\epsilon\left(e^\mu - 1\right)} \int_{0}^{\beta \epsilon} d\tau ~ \tau \exp\left(\frac{\mu}{\beta\epsilon}\tau\right) \left(\frac{1}{\tau}\partial_\tau(\tau \partial_\tau) - \frac{1}{\tau^2}\right)^k \left(\frac{\tanh^2 \tau}{\tau} \right).
	\end{aligned}
\end{equation}
Now we conclude that
\begin{align}
	c_{2k} &\,=\frac{2\pi}{\lambda_0^{2k}}\left[D f^{(2k)}(\beta, \epsilon, \mu) - J g^{(2k)}(\beta, \epsilon)\right],
\end{align}
where
\begin{gather}
	f^{(2k)}(\beta, \epsilon, \mu) = - \frac{\beta^{2k - 1} \mu}{2^{k}\epsilon\left(e^\mu - 1\right)} \int_{0}^{\beta \epsilon} d\tau ~ \tau \exp\left(\frac{\mu}{\beta\epsilon}\tau\right)\left(\frac{1}{\tau}\partial_\tau(\tau \partial_\tau) - \frac{1}{\tau^2}\right)^k \frac{\tanh^2 \tau}{\tau},\\
	g^{(2k)}(\beta, \epsilon) = -\frac{\beta^{2k - 2}}{2^{k + 1}} \left[\tau \partial_\tau \left(\frac{1}{\tau}\partial_\tau(\tau \partial_\tau)\right)^{k - 1}\left((\beta^2 + 1)\,\text{sech}^2\tau + \frac{\beta^2}{\tau^2} \tanh^2\tau\right)\right]\Biggr|_{\tau = 0}^{\beta \epsilon}.
\end{gather}

\renewcommand{\theequation}{B\arabic{equation}}
\setcounter{equation}{0}

\section{Analytic solution of the anharmonic vortex oscillation}\label{ApB}
In this section, we solve the nonlinear ordinary differential equation given in Eq.~\eqref{ode}. Let $\psi = \mathcal{R}e^{i\phi}$, where $\mathcal{R} = \mathcal{R}(t)$ and $\phi = \phi(t)$ are real-valued functions. Then Eq.~\eqref{ode} becomes
\begin{gather}
	\dot{\mathcal{R}} + i\dot{\phi} \mathcal{R}  -i \omega \mathcal{R} -i \lambda \mathcal{R}^3 = 0.
\end{gather}
Separating the real and imaginary parts, we obtain
\begin{gather}
	\left(\dot{\mathcal{R}} + \omega_i \mathcal{R} + \lambda_i \mathcal{R}^3\right) + i \left(\dot{\phi} - \omega_r - \lambda_r \mathcal{R}^2\right) \mathcal{R} = 0\,,
\end{gather}
which yields the couple of equations:
\begin{gather}\label{couple}
	\begin{cases}
		\dot{\mathcal{R}} + \omega_i \mathcal{R} + \lambda_i \mathcal{R}^3 = 0\\
		\dot{\phi} - \omega_r - \lambda_r \mathcal{R}^2 = 0
	\end{cases}.
\end{gather}
The first equation can be solved by the seperation of variables, with $\mathcal{R}_0 = \mathcal{R}(t = 0)$:
\begin{gather}
	\mathcal{R}(t) = \frac{\mathcal{R}_0 \exp(- \omega_i t)}{\sqrt{1 + \frac{\lambda_i \mathcal{R}_0^2}{\omega_i} \left(1 - \exp(-2\omega_i t)\right)}}\,,
\end{gather}
and the solution of the second equation follows from direct integration:
\begin{align}
	\phi(t) = \phi_0 + \omega_r t + \frac{\lambda_r}{2\lambda_i} \ln \left(1 + \frac{\lambda_i \mathcal{R}_0^2}{\omega_i}\left( 1 - \exp(-2 \omega_i t)\right)\right),
\end{align}
where $\phi_0 \equiv \phi(t = 0)$. In summary, the full solution is given by
\begin{align}
	\psi(t) = \psi_0 e^{i \omega t} \frac{\exp\left(i\frac{\lambda_r}{\lambda_i}\ln\sqrt{1 + \frac{\lambda_i \mathcal{R}_0^2}{\omega_i}\left(1 - e^{-2\omega_i t}\right)}\right)}{\sqrt{1 + \frac{\lambda_i \mathcal{R}_0^2}{\omega_i}\left(1 - e^{-2\omega_i t}\right)}}\,,
\end{align}
where $\psi_0 =\psi(t = 0)$ and $\mathcal{R}_0 = |\psi_0|$. Furthermore, using
\begin{align}
	\sigma(t) = \frac{1}{\lambda_i} \ln \sqrt{1 + \frac{\lambda_i \mathcal{R}_0^2}{\omega_i}\left(1 - e^{-2\omega_i t}\right)}\,,
\end{align}
we have
\begin{equation}
	\begin{aligned}
		\psi(t) &\,= \psi_0 e^{i \omega t} \frac{\exp\left[i \lambda_r \sigma(t)\right]}{\exp\left[\lambda_i \sigma(t)\right]} = \psi_0 \exp\left[i \omega t + i \lambda \sigma(t)\right]\,.
	\end{aligned}
\end{equation}

\renewcommand{\theequation}{C\arabic{equation}}
\setcounter{equation}{0}

\section{Details on the driven oscillation of the anharmonic vortex oscillation}\label{ApC}
In this section, we present the details of the driven oscillation under oscillating external torques, such as a external magnetic field $\mathbf{B}(t) = B_x \sin(\omega_d t)\hat{\mathbf{x}}$ or the adiabatic Zhang--Li spin-transfer torque~\cite{Li_2004} with a current $\mathbf{j}(t) = j_x \sin(\omega_d t) \hat{\mathbf{x}}$.

In the case of a magnetic field, the right-hand side of Eq.~\eqref{thiele} acquires an additional term given by
\begin{equation}
	\begin{gathered}
		\gamma \int d^2x 
		\begin{pmatrix}
			(\partial_X \mathbf{m}) \cdot \mathbf{B}\\
			(\partial_Y \mathbf{m}) \cdot \mathbf{B}
		\end{pmatrix}
		= \pi \gamma R\, \text{sgn}(h) \tanh(\beta \epsilon) B_x \sin(\omega_d t)
		\begin{pmatrix}
			0\\
			1
		\end{pmatrix},
	\end{gathered}
\end{equation}
and in the case of adiabatic Zhang-Li spin-transfer torque, where the torque is given by $\bm{\tau}_\text{ZL} = \frac{\gamma \hbar P}{2eM_\text{sat}a^3}\mathbf{j} \cdot \nabla \mathbf{m}$,
\begin{equation}
	\begin{gathered}
		-\int d^2x
		\begin{pmatrix}
			\mathbf{m} \cdot \left[\left(\partial_X \mathbf{m}\right) \times \bm{\tau}\right]\\
			\mathbf{m} \cdot \left[\left(\partial_Y \mathbf{m}\right) \times \bm{\tau}\right]
		\end{pmatrix}
		=
		-\frac{\pi \gamma \hbar p P}{eM_\text{sat} a^3} j_x \sin(\omega_d t)
		\begin{pmatrix}
			0\\
			1
		\end{pmatrix}.
	\end{gathered}
\end{equation}
In both cases, the vortex dynamics can be described by the equation of motion
\begin{gather}\label{eq_full}
	\dot{\psi} = i \omega \psi + i \lambda |\psi|^2 \psi + F_0 \sin(\omega_d t)\,,
\end{gather}
as in Eq.~\eqref{driven}, where
\begin{empheq}[left={F_0 = \empheqlbrace}]{equation}
	\begin{aligned}
		&\frac{\gamma R\,\text{sgn}(h) \tanh(\beta \epsilon)}{2[\alpha I(\beta, \epsilon) + i p]}B_x&&(\text{driven by magnetic field})\\
		&\frac{\gamma \hbar p P}{2e M_\text{sat} a^3[\alpha I(\beta, \epsilon) + i p]}j_x&&(\text{driven by Zhang-Li STT})
	\end{aligned}.
\end{empheq}

Using the method of harmonic balance~\cite{10.1093/oso/9780199208241.001.0001} with the ansatz
\begin{align}\label{ansatz}
	\psi(t) = \mathcal{R}(t) e^{i(\omega_d t + \phi_0)},
\end{align}
Eq.~\eqref{eq_full} becomes
\begin{gather}
	\dot{\mathcal{R}} e^{i(\omega_d t + \phi_0)} + i \omega_d \mathcal{R} e^{i(\omega_d t + \phi_0)} = i \omega \mathcal{R} e^{i(\omega_d t + \phi_0)} + i \lambda \mathcal{R}^3 e^{i(\omega_d t + \phi_0)} + F_0 \sin(\omega_d t)\\
	\Longrightarrow i\left[-i\dot{\mathcal{R}} + \omega_d - \left(\omega + \lambda \mathcal{R}^2\right)\right]\mathcal{R} e^{i\omega_d t} = F_0 e^{-i \phi_0} \sin(\omega_d t)\,.
\end{gather}
Taking the modulus square yields
\begin{align}\label{response_ns}
	\left|-i\dot{\mathcal{R}} + \omega_d - \left(\omega + \lambda \mathcal{R}^2\right)\right|^2 \mathcal{R}^2 = \frac{|F_0|^2}{2} \left[1 - \cos(2\omega_d t)\right].
\end{align}
Assuming a quasi-stationary radius of gyration ($\dot{\mathcal{R}} \approx 0$) and neglecting the higher-order oscillatory term at frequency $2\omega_d$, we obtain Eq.~\eqref{response}.

To analyze stability, we linearize Eq.~\eqref{eq_full} around the solution $\psi$ and consider a deviation $\delta \psi = \delta X + i\delta Y$ of the motion:
\begin{align}
	\delta \dot{\psi} = i\omega \delta \psi + 2 i \lambda |\psi|^2 \delta \psi  + i\lambda \psi^2 \delta \psi^\ast = i\omega \delta \psi + i\lambda \mathcal{R}^2 e^{2i(\omega_d t + \phi_0)} \delta \psi^\ast + 2i\lambda \mathcal{R}^2 \delta \psi\,,
\end{align}
which yields the real number form
\begin{align}
	\frac{d}{dt}
	\begin{pmatrix}
		\delta X\\
		\delta Y
	\end{pmatrix}
	=
	\begin{pmatrix}
		-\text{Im}\left[\omega + \lambda \mathcal{R}^2 e^{2i(\omega_d t + \phi_0)} + 2\lambda \mathcal{R}^2\right]&-\text{Re}\left[\omega - \lambda \mathcal{R}^2 e^{2i(\omega_d t + \phi_0)} + 2\lambda \mathcal{R}^2\right]\\
		\text{Re}\left[\omega + \lambda \mathcal{R}^2 e^{2i(\omega_d t + \phi_0)} + 2\lambda \mathcal{R}^2\right]&-\text{Im}\left[\omega - \lambda \mathcal{R}^2 e^{2i(\omega_d t + \phi_0)} + 2\lambda \mathcal{R}^2\right]
	\end{pmatrix}
	\begin{pmatrix}
		\delta X\\
		\delta Y
	\end{pmatrix}.
\end{align}
According to Floquet-Lyapunov theory~\cite{10.1093/oso/9780199208241.001.0001}, stability is determined from the solution of
\begin{align}
	\dot{M}(t) =
	\begin{pmatrix}
		-\text{Im}\left[\omega + \lambda \mathcal{R}^2 e^{2i(\omega_d t + \phi_0)} + 2\lambda \mathcal{R}^2\right]&-\text{Re}\left[\omega - \lambda \mathcal{R}^2 e^{2i(\omega_d t + \phi_0)} + 2\lambda \mathcal{R}^2\right]\\
		\text{Re}\left[\omega + \lambda \mathcal{R}^2 e^{2i(\omega_d t + \phi_0)} + 2\lambda \mathcal{R}^2\right]&-\text{Im}\left[\omega - \lambda \mathcal{R}^2 e^{2i(\omega_d t + \phi_0)} + 2\lambda \mathcal{R}^2\right]
	\end{pmatrix}
	M(t)
\end{align}
where $M(t)$ is a $2\times 2$ matrix with the initial condition
\begin{align}
	M(t_0) = 
	\begin{pmatrix}
		1&0\\
		0&1
	\end{pmatrix}.
\end{align}
Let $\mu_1$ and $\mu_2$ be the eigenvalues of the monodromy matrix $M(t_0 + 2\pi / \omega_d)$. The driven oscillation is dynamically stable when $|\mu_1| \le 1$ and $|\mu_2| \le 1$; otherwise, it becomes unstable~\cite{10.1093/oso/9780199208241.001.0001}.

\twocolumngrid
\bibliography{manuscript.bbl}

\end{document}